\documentclass[showkeys,superscriptaddress,floatfix,prd,10pt,aps,twocolumn]{revtex4-2}
\usepackage{graphicx,epstopdf}
\usepackage[caption=false]{subfig}
\usepackage{appendix}
\usepackage[T1]{fontenc}
\usepackage{lmodern}
\usepackage[dvipsnames,x11names]{xcolor}
\usepackage[colorlinks=true,linkcolor=ForestGreen,citecolor=ForestGreen,urlcolor=NavyBlue]{hyperref}
\usepackage[sort&compress]{natbib}
\usepackage{lipsum}
\usepackage{morefloats}
\usepackage[pdf]{pstricks}
\usepackage{amsmath}
\usepackage{amssymb}
\usepackage{amsfonts}
\usepackage{rotating}
\usepackage{cancel}
\usepackage{mathtools}
\usepackage{bbm}
\usepackage{dsfont}
\usepackage{bbold}
\usepackage{multirow}
\usepackage{ulem}
\usepackage{physics}
\usepackage{orcidlink}
\usepackage{colortbl}

\def\pslash{p\!\!\!\slash }
\def\qslash{q\!\!\!\slash }
\def\xslash{x\!\!\!\slash }
\def\eslash{\varepsilon\!\!\!\slash }

\def\vel{\left|}
\def\ver{\right|}

\begin{document}

\title{Electromagnetic properties of the $\Sigma_{c}(2800)^+$ and $\Lambda_c(2940)^+$ states via light-cone QCD}

\author{Ula\c{s}~\"{O}zdem\orcidlink{0000-0002-1907-2894}}%
\email[]{ulasozdem@aydin.edu.tr }
\affiliation{ Health Services Vocational School of Higher Education, Istanbul Aydin University, Sefakoy-Kucukcekmece, 34295 Istanbul, T\"{u}rkiye}

\date{\today}
 
\begin{abstract}
The $\Sigma_{c}(2800)^+$ and $\Lambda_c(2940)^+$ states are among the most interesting and intriguing particles whose internal structures have not yet been elucidated. Inspired by this, the magnetic dipole moments of the $\Sigma_{c}(2800)^+$ and $\Lambda_c(2940)^+$  states with quantum numbers $J^P = \frac{1}{2}^-$ and $J^P = \frac{3}{2}^-$, respectively, are analyzed in the framework of QCD light-cone sum rules, assuming that they have a molecule composed of a nucleon and a $D^{(*)}$ meson. The magnetic dipole moments are obtained as  $\mu_{\Sigma_c^{+}}=0.26 \pm 0.05~\mu_N$ and   $\mu_{\Lambda_c^{+}}=-0.31 \pm 0.04~\mu_N$. The magnetic dipole moment is the leading-order response of a bound system to a weak external magnetic field. It therefore offers an excellent platform to explore the inner organization of hadrons governed by the quark-gluon dynamics of QCD. Comparison of the findings of this analysis with future experimental results on the magnetic dipole moments of the $\Sigma_{c}(2800)^+$ and $\Lambda_c(2940)^+$  states may shed light on the nature and internal organization of these states. The electric quadrupole and magnetic octupole moments of the $\Lambda_c(2940)^+$ states have also been calculated, and these values are determined to be $\mathcal Q_{\Lambda_c^+} = (0.65 \pm 0.25) \times 10^{-3}$~fm$^2$ and $ \mathcal O_{\Lambda_c^+} = -(0.38 \pm 0.10) \times 10^{-3}$~fm$^3$, respectively. The values of the electric quadrupole and magnetic octupole moments show a non-spherical charge distribution. We hope that our estimates of the electromagnetic properties of the $\Sigma_{c}(2800)^+$ and $\Lambda_c(2940)^+$ states, together with the results of other theoretical studies of the spectroscopic parameters of these states, will be useful for their search in future experiments and will help us to define the exact internal structures of these states.
\end{abstract}

\maketitle

\section{Introduction}
 
In the past few years, many excited baryon states have been discovered experimentally. Elucidating and understanding the internal structure of excited hadronic states is an important problem in the field of nonperturbative QCD. The discovery of a large number of new heavy baryons at experimental facilities requires their theoretical identification and classification \cite{Guo:2017jvc,Olsen:2017bmm,Brambilla:2019esw,Chen:2022asf}. 
Theoretical investigations include the analysis of baryons containing a single heavy quark, which provides an excellent laboratory to study the dynamics of a light diquark in the environment of a heavy quark, allowing the predictions of different theoretical approaches to be tested and to improve the understanding of the non-perturbative nature of QCD.
 Significant experimental progress has been made in the field of singly heavy baryons in recent years 
 \cite{ARGUS:1993vtm, CLEO:1994oxm, ARGUS:1997snv, E687:1993bax, CLEO:2000mbh, BaBar:2006itc, Belle:2006xni, LHCb:2017jym, Ammosov:1993pi, CLEO:1996czm, Belle:2004zjl, CLEO:1995amh, CLEO:1996zcj,E687:1998dwp, CLEO:2000ibb, CLEO:1999msf, LHCb:2020iby, BaBar:2007xtc, Belle:2006edu, BaBar:2007zjt, BaBar:2006pve, LHCb:2017uwr, LHCb:2012kxf, LHCb:2020lzx, LHCb:2019soc, LHCb:2018haf, CMS:2021rvl,  LHCb:2018vuc,  LHCb:2020xpu, LHCb:2020tqd}. More experimental research is needed because the evidence for the existence of some of these states is weak and the quantum numbers are not well determined. Therefore, their study is an active field in both experimental and theoretical research. 
 
In 2005, the Belle Collaboration tentatively identified the $\Sigma_c(2800)^{++}$, $\Sigma_c(2800)^{+}$  and   $\Sigma_c(2800)^{0}$ as isospin triplet states in the $\Lambda_c^+\pi^{+}$, $\Lambda_c^+\pi^{0}$ and $\Lambda_c^+\pi^{-}$ mass spectrum, respectively \cite{Belle:2004zjl}. The measured mass differences and decay widths are listed as follows,

 \begin{align}
 &M_{\Sigma_c(2800)^{++}} - M_{\Lambda_c^+}= 514.5^{+3.4}_{-3.1}{}^{+2.8}_{-4.9} \,\mbox{MeV}\,,\nonumber\\
 &\Gamma_{\Sigma_c(2800)^{++}}=75^{+18}_{-13}{}^{+12}_{-11}\mbox{ MeV}\,, \\
 \nonumber\\
 & M_{\Sigma_c(2800)^{+}} - M_{\Lambda_c^+}=505.4^{+5.8}_{-4.6}{}^{+12.4}_{-2.0} \,\mbox{MeV}\,,\nonumber\\
 & \Gamma_{\Sigma_c(2800)^{+}}=62^{+37}_{-23}{}^{+52}_{-38} \mbox{ MeV}\,,\\
  \nonumber\\
& M_{\Sigma_c(2800)^{0}} - M_{\Lambda_c^+}=515.4^{+3.2}_{-3.1}{}^{+2.1}_{-6.0}\, \mbox{MeV}\,,\nonumber\\
& \Gamma_{\Sigma_c(2800)^{0}}=61^{+18}_{-13}{}^{+22}_{-13} \mbox{ MeV}\,.
\end{align}
 
The $\Sigma_c(2800)^{0}$ state is probably confirmed by the BaBar Collaboration \cite{BaBar:2008get} and; the mass and decay width are determined to be $ M_{\Sigma_c(2800)^0} = 2846\pm 8\pm 10 \mbox{ MeV}$, $\Gamma_{\Sigma_c(2800)^0} = 86{}^{+33}_{-22}\pm 7\mbox{ MeV}$. However, the measured width of this state is consistent with the Belle data, the mass value is about $50 \mbox{ MeV}$ larger and somewhat inconsistent with the previous measurement. If the particle observed by the two experimental collaborations is indeed in the same state, then the discrepancy in the measured masses must be resolved. 
The quantum numbers of these states also remain to be determined.  

In 2007, the BaBar Collaboration searched for charmed baryons in the $D^0p$ invariant mass spectrum and found the $\Lambda_c(2940)^+$ \cite{BaBar:2006itc}. The Belle Collaboration then confirmed the $\Lambda_c(2940)^+$ in decay mode $\Lambda_c(2940) \to \Sigma_c(2455) \pi$ \cite{Belle:2006xni}. In 2017, the LHCb collaboration measured the amplitude analysis of the $\Lambda_b^0  \to D^0p \pi^-$ process to determine the spin-parity of the $\Lambda(2940)^+$ state, and the most likely spin-parity assignment for this state is $J^P=\frac{3}{2}^-$ \cite{LHCb:2017jym}. The masses and decay widths measured for this state by the three collaborations are given below,

\begin{align}
 &M_{\Lambda_c(2940)^+} = 2939.8 \pm1.3 \pm1.0 \mbox{ MeV}\,,\nonumber\\
 &\Gamma_{\Lambda_c(2940)} = 17.5\pm5.0\pm5.9 \mbox{ MeV}\, 
 ~\mbox{(BaBar)}\,, 
 \\
 \nonumber \\
& M_{\Lambda_c(2940)^+} = 2938.0 \pm1.3 ^{+2.0}_{-4.0} \mbox{ MeV}\,,\nonumber\\
&\Gamma_{\Lambda_c(2940)} = 13{}^{+8}_{-5}{}^{+27}_{-7} \mbox{ MeV}\, 
~\mbox{(Belle)}, 
\\
 \nonumber \\
 &M_{\Lambda_c(2940)^+} = 2944.8 ^{+3.5}_{-2.5} \pm0.4 {}^{+0.1}_{-4.6} \mbox{ MeV}\,,\,\nonumber\\
 &\Gamma_{\Lambda_c(2940)} = 27.7^{+8.2}_{-6.0}\pm0.9{}^{+5.2}_{-10.4} \mbox{ MeV} \,
 ~\mbox{(LHCb)}\,.
\end{align}
 
Following these experimental observations, understanding the internal structures and quantum numbers of the $\Sigma_{c}(2800)^+$ and $\Lambda_c(2940)^+$  states has been an important topic of research among theorists. 
To elucidate their underlying structure, they were studied in standard baryon \cite{Wang:2017kfr, Yang:2021lce, Jia:2019bkr, Zhou:2023wrf, Wang:2022dmw, Guo:2019ytq, Lu:2018utx, Luo:2019qkm, Gong:2021jkb} and molecular pentaquark states 
\cite{Sakai:2020psu, Zhao:2016zhf, Wang:2018jaj, Zhang:2019vqe, Yan:2023ttx, Zhang:2012jk, Wang:2020dhf, Xin:2023gkf, He:2006is,Yan:2022nxp, Zhang:2022pxc, Huang:2016ygf, Wang:2015rda, Dong:2011ys, He:2010zq, Dong:2010xv, Dong:2009tg, Ortega:2012cx, Dong:2014ksa}. The $\Sigma_{c}(2800)^+$ and $\Lambda_c(2940)^+$  states are still not yet fully understood, despite the considerable efforts that have been devoted to their study.  All of these works, done with alternative structures however with predictions that are within error of the experimental observations, indicate the need for more for further study of the $\Sigma_{c}(2800)^+$ and $\Lambda_c(2940)^+$ states. Therefore, it is time for further efforts in the study of these states to have a better understanding of their properties.

If one examines the studies in the literature listed above, it can be seen that almost all the calculations are aimed at calculating the spectroscopic parameters of these states, and it is easy to see that the spectroscopic parameters alone are not sufficient to elucidate the controversial nature of these states. Hence, it is obvious that further analyses such as the electromagnetic form factors, the weak decays, and so on are needed to shed light on the internal structure of these states. 
The $\Sigma_{c}(2800)^+$ and $\Lambda_c(2940)^+$  states are close to the $ND$ and $ND^*$ thresholds, inspiring us that they might be the $ND$ and $ND^*$ molecular states. Then it is worthwhile to study the $ND$ and $ND^*$ interactions with various methods to further understand the nature of the $\Sigma_{c}(2800)^+$ and $\Lambda_c(2940)^+$  states. 
Based on this assumption, we consider the states $\Sigma_{c}(2800)^+$ and $\Lambda_c(2940)^+$  to be $ND$ and $ND^*$ bound states: 
$\Sigma_c^{+}=(\mid D^0 p \rangle \, + \mid D^+ n \rangle )/ \sqrt{2}$,  
and $\Lambda_c^{+}=(\mid D^{*0} p \rangle \, - \mid D^{*+} n \rangle )/ \sqrt{2}$; and we explore the nature of these states in the framework of QCD light-cone sum rules. 
It is worth noting that, as mentioned above, the spin-parity quantum numbers of these states have not been experimentally determined. However, when considering the molecular state, studies in the literature have shown that the $J^P = \frac{1}{2}^-$ assignment for the $\Sigma_{c}(2800)^+$ state and the $J^P = \frac{3}{2}^-$ assignment for the $\Lambda_c(2940)^+$ state are more consistent with the experimental data. Therefore, in this study we assume that the $\Sigma_{c}(2800)^+$ and $\Lambda_c(2940)^+$ states have spin-parity quantum numbers $J^P = \frac{1}{2}^-$ and $J^P = \frac{3}{2}^-$, respectively. 

In the present paper, we proceed as follows. We derive QCD light-cone sum rules for molecular states in Sect. \ref{secII}, with similar procedures as our previous studies on baryons~\cite{Ozdem:2018uue,Ozdem:2019zis,Ozdem:2023cri} and molecular states~\cite{Ozdem:2021ugy,Ozdem:2023htj,Ozdem:2022kei,Ozdem:2021vry,Ozdem:2021hmk,Ozdem:2022vip}. In Sect. \ref{secIII}, we illustrate our numerical results and discussions, and the last section is devoted to the summary and concluding remarks.
 The analytical expressions obtained for the magnetic dipole moment results of the $\Sigma_{c}^{+}$ state, electromagnetic multipole moments of the $\Lambda_c(2940)^+$ state, and the photon DAs are presented in Appendix \ref{appenda}, \ref{appendb} and \ref{appendc}, respectively.  

  
\section{QCD light-cone sum rules for the electromagnetic properties of the $\Sigma_{c}(2800)^+$ and $\Lambda_c(2940)^+$ states}\label{secII}

As is well known, the QCD light-cone sum rules have been successfully applied to the calculation of masses, decay constants, form factors, magnetic dipole moments, etc. of standard hadrons, and are a powerful technique for the study of exotic hadron properties. The correlation function is evaluated both in terms of hadrons (the hadronic side) and in terms of quark-gluon degrees of freedom (the QCD side) according to the QCD light cone sum rules technique.   Then, by equating these two different descriptions of the correlation function by quark-hadron duality, the physical quantities, i.e. the magnetic dipole moments, are calculated~\cite{Chernyak:1990ag,Braun:1988qv,Balitsky:1989ry}.

\subsection{Magnetic dipole moments of the $\Sigma_{c}(2800)^+$ state} 

 The first step to analyze magnetic dipole moments with the QCD light-cone sum rules is to write the following correlation function,
 \begin{eqnarray} \label{edmn01}
\Pi(p,q)&=&i\int d^4x e^{ip \cdot x} \langle0|T\left\{J^{\Sigma_{c}^+}(x)\bar{J}^{\Sigma_{c}^+}(0)\right\}|0\rangle _\gamma \, , 
\end{eqnarray}
where $T$ is the time ordered product, sub-indice $\gamma$ is the external electromagnetic field. $J(x)$ is the interpolating current of the $\Sigma_{c}^+$ state and is required to continue the analysis. This interpolating current is written as follows by the quantum numbers $I(J^P) = 1(1/2^-)$:
\begin{align}\label{curpcs1}
J^{\Sigma_{c}^+}(x)& =\frac{1}{\sqrt{2}}\Big \{\mid  D^0 p \rangle \, + \mid  D^+ n \rangle  \Big \}\nonumber\\
&
=\frac{1}{\sqrt{2}} \Big \{ \big[\bar u^d(x)i \gamma_5 c^d(x)\big]\big[\varepsilon^{abc} u^{a^T}(x)C\gamma_\mu d^b(x) \nonumber\\
& \times  \gamma^\mu\gamma_5 u^c(x)\big] - \big[\bar d^d(x)i \gamma_5 c^d(x)\big]  
 \big[\varepsilon^{abc} u^{a^T}(x)\nonumber\\
&
 \times C\gamma_\mu d^b(x) 
 \gamma^\mu\gamma_5 d^c(x)\big] \Big\} \, , 
 \end{align}
where $a$, $b$, $c$ and  $d$ are color indices and the $C$ is the charge conjugation operator.

Let us start by obtaining the correlation function from the hadronic side.
We insert a full set of intermediate state $\Sigma_{c}$ with the same quantum numbers as the interpolating currents into the correlation function to get the hadronic representation of the correlation function. This gives us the following expression
 \begin{align}\label{edmn02}
\Pi^{Had}(p,q)&=\frac{\langle0\mid J^{\Sigma_{c}^+}(x) \mid
{\Sigma_{c}^+}(p, s) \rangle}{[p^{2}-m_{\Sigma_{c}^+}^{2}]}
\nonumber\\
& \times 
\langle {\Sigma_{c}^+}(p, s)\mid
{\Sigma_{c}^+}(p+q, s)\rangle_\gamma 
\nonumber\\
& \times 
\frac{\langle {\Sigma_{c}^+}(p+q, s)\mid
\bar J^{\Sigma_{c}^+}(0) \mid 0\rangle}{[(p+q)^{2}-m_{\Sigma_{c}^+}^{2}]}+ \cdots 
\end{align}

The matrix elements in Eq.~(\ref{edmn02}) can be written in terms of hadronic parameters and Lorentz invariant form factors in the following way:
%
\begin{align} 
\langle0\mid J^{\Sigma_{c}^+}(x)\mid {\Sigma_{c}^+}(p, s)\rangle=&\lambda_{\Sigma_{c}^+} \gamma_5 \, u(p,s),\label{edmn04}\\
\langle {\Sigma_{c}^+}(p+q, s)\mid\bar J^{\Sigma_{c}^+}(0)\mid 0\rangle=&\lambda_{\Sigma_{c}^+} \gamma_5 \, \bar u(p+q,s)\label{edmn004}
,\\
\langle {\Sigma_{c}^+}(p, s)\mid {\Sigma_{c}^+}(p+q, s)\rangle_\gamma &=\varepsilon^\mu\,\bar u(p, s)\Big[\big[F_1(q^2)
\nonumber\\
&
+F_2(q^2)\big] \gamma_\mu +F_2(q^2)
\nonumber\\
& \times 
\frac{(2p+q)_\mu}{2 m_{\Sigma_{c}^+}}\Big]\,u(p+q, s), \label{edmn005}
\end{align}
where the $u(p,s)$, $ u(p+q,s)$ and $\lambda_{{\Sigma_c^+}}$ are the spinors and residue of the $\Sigma_c^+$ state, respectively.

Substituting Eqs. (\ref{edmn04})-(\ref{edmn005}) in Eq. (\ref{edmn02}) and doing some calculations, we get the following result for the hadronic side,
\begin{align}
\label{edmn05}
\Pi^{Had}(p,q)=&\lambda^2_{\Sigma_{c}^+}\gamma_5 \frac{\Big(\pslash+m_{\Sigma_{c}^+} \Big)}{[p^{2}-m_{{\Sigma_{c}^+}}^{2}]}\varepsilon^\mu \Bigg[\big[F_1(q^2) %
+F_2(q^2)\big] \gamma_\mu
\nonumber\\
& 
+F_2(q^2)\, \frac{(2p+q)_\mu}{2 m_{\Sigma_{c}^+}}\Bigg]  \gamma_5 
\frac{\Big(\pslash+\qslash+m_{\Sigma_{c}^+}\Big)}{[(p+q)^{2}-m_{{\Sigma_{c}^+}}^{2}]}. 
\end{align}
To obtain the above expression, summing over the spins of $\Sigma_{c}^+$ state, $\sum_s u(p,s)\bar u(p,s)=\pslash+m_{\Sigma_{c}^+}$ and $\sum_s u(p+q,s)\bar u(p+q,s)=(\pslash+\qslash)+m_{\Sigma_{c}^+}$ have also been performed.

 The magnetic dipole moments of hadrons are related to their magnetic form factors; more precisely, the magnetic dipole moments are equal to the magnetic form factor at zero momentum squared. The magnetic form factors ($F_M(q^2)$), which are more directly accessible in experiments, are defined by the form factors $F_1(q^2)$ and $F_2(q^2)$
\begin{align}
\label{edmn07}
&F_M(q^2) = F_1(q^2) + F_2(q^2).
\end{align}
 Since we are dealing with a real photon, we can define the magnetic form factor $F_M (q^2 = 0)$ in terms of the magnetic dipole moment $\mu_{\Sigma_{c}^+}$:
\begin{align}
\label{edmn08}
&\mu_{\Sigma_{c}^+} = \frac{ e}{2\, m_{\Sigma_{c}^+}} \,F_M(q^2 = 0).
\end{align}

Now we are ready to start the second step, which is the calculation of the correlation function using the QCD parameters. 
By explicitly using the interpolating currents in the correlation function, the second representation of the correlation function, the QCD side, is obtained.  Then, all of the quark fields are contracted according to Wick's theorem, and the desired results are obtained. As a result of the procedures described above, we obtain the following:
\label{QCD1}
%
\begin{align}
\label{QCD2}
\Pi^{QCD}_{\Sigma_c^+}(p,q)&=- \frac{i}{2}\varepsilon^{abc} \varepsilon^{a^{\prime}b^{\prime}c^{\prime}}\, \int d^4x \, e^{ip\cdot x} \langle 0\mid \Big\{ 
\nonumber\\
& 
\, \mbox{Tr}\Big[\gamma_5 S_{c}^{dd^\prime}(x) \gamma_5  S_{u}^{d^\prime d}(-x)\Big]  
\mbox{Tr}\Big[\gamma_{\alpha} S_d^{bb^\prime}(x) \gamma_{\beta}  
\nonumber\\
& \times  \widetilde S_{u}^{aa^\prime}(x)\Big]
(\gamma^{\alpha}\gamma_5 S_{u}^{cc^\prime}(x) \gamma_5  \gamma^{\beta})
 \nonumber\\
&     
- \mbox{Tr}\Big[\gamma_5 S_{c}^{dd^\prime}(x) \gamma_5  S_{u}^{d^\prime d}(-x)\Big]   
 \mbox{Tr}\Big[\gamma^{\alpha}\gamma_5 S_{u}^{ca^\prime}(x)
 \nonumber\\
& \times  \gamma_{\beta} \widetilde S_d^{bb^\prime}(x) \gamma_{\alpha}  
 S_{u}^{ac^\prime}(x)  \gamma_5  \gamma^{\beta}\Big]  \nonumber\\
&
  + \mbox{Tr}\Big[\gamma_5 S_{c}^{dd^\prime}(x) \gamma_5  S_{d}^{d^\prime d}(-x)\Big]  
\mbox{Tr}\Big[\gamma_{\alpha} S_d^{bb^\prime}(x) 
\nonumber\\
& \times \gamma_{\beta} \widetilde S_{u}^{aa^\prime}(x)\Big]
(\gamma^{\alpha}\gamma_5 S_{d}^{cc^\prime}(x) \gamma_5  \gamma^{\beta})
 \nonumber\\
&  -  \mbox{Tr}\Big[\gamma_5 S_{c}^{dd^\prime}(x) \gamma_5  S_{d}^{d^\prime d}(-x)\Big]   
 \mbox{Tr}\Big[\gamma^{\alpha}\gamma_5 S_{d}^{cb^\prime}(x) 
 \nonumber\\
& \times \gamma_{\beta} \widetilde S_u^{aa^\prime}(x) \gamma_{\alpha}  
 S_{d}^{bc^\prime}(x)  \gamma_5  \gamma^{\beta}\Big] 
 \Big\}
\mid 0 \rangle _\gamma \,,
\end{align}
%
where   
$\widetilde{S}_{c(q)}^{ij}(x)=CS_{c(q)}^{ij\mathrm{T}}(x)C$ and,
 $S_{c}(x)$ and $S_{q}(x)$ 
 are the full propagators of the heavy and light quarks, which can be written as~\cite{Yang:1993bp, Belyaev:1985wza}
\begin{align}
\label{edmn13}
S_{q}(x)&= S_q^{free} 
- \frac{\langle \bar qq \rangle }{12} \Big(1-i\frac{m_{q} \xslash}{4}   \Big)
- \frac{ \langle \bar qq \rangle }{192}
m_0^2 x^2  \Big(1
\nonumber\\
&  -i\frac{m_{q} \xslash}{6}   \Big)
-\frac {i g_s }{32 \pi^2 x^2} ~G^{\mu \nu} (x) 
\Big[\rlap/{x} 
\sigma_{\mu \nu} +  \sigma_{\mu \nu} \rlap/{x}
 \Big],
\end{align}%
\begin{align}
\label{edmn14}
S_{c}(x)&=S_c^{free}
-\frac{g_{s}m_{c}}{16\pi ^{2}} \int_0^1 dv\, G^{\mu \nu }(vx)\Bigg[ (\sigma _{\mu \nu }{\xslash}
  +{\xslash}\sigma _{\mu \nu }) 
  \nonumber\\
& \times 
  \frac{K_{1}\Big( m_{c}\sqrt{-x^{2}}\Big) }{\sqrt{-x^{2}}}
 +2\sigma_{\mu \nu }K_{0}\Big( m_{c}\sqrt{-x^{2}}\Big)\Bigg],
\end{align}%
and 
\begin{align}
 S_q^{free}&=\frac{1}{2 \pi x^2}\Big(i \frac{\xslash}{x^2}- \frac{m_q}{2}\Big),\\
 S_c^{free}&=\frac{m_{c}^{2}}{4 \pi^{2}} \Bigg[ \frac{K_{1}\Big(m_{c}\sqrt{-x^{2}}\Big) }{\sqrt{-x^{2}}}
+i\frac{{\xslash}~K_{2}\Big( m_{c}\sqrt{-x^{2}}\Big)}
{(\sqrt{-x^{2}})^{2}}\Bigg],
\end{align}
where $\langle \bar qq \rangle$ is light-quark condensate, $G^{\mu\nu}$ is the gluon field strength tensor, $v$ is line variable and  $K_i$'s are modified Bessel functions of the second kind, respectively.    

The correlation functions in Eq.~(\ref{QCD2}) receive both perturbative contributions, that is, when a photon interacts with light/heavy quarks at short-distance and non-perturbative contributions, that is when a photon interacts with light quarks at large-distance. 
In the case of perturbative contributions, one of the light propagators or one of the heavy propagators of the quarks interacting perturbatively with the photon is replaced by the following
\begin{align}
\label{free}
S^{free}(x) \rightarrow \int d^4y\, S^{free} (x-z)\,\rlap/{\!A}(z)\, S^{free} (z)\,,
\end{align}
and the rest of the propagators in Eq.~(\ref{QCD2}) are considered full quark propagators, which include perturbative and non-perturbative contributions.   The overall perturbative contribution is obtained by replacing the perturbatively interacting quark propagator with the photon as described above, and by replacing the surviving propagators with their free parts. Here we use $ A_\mu(z)=\frac{i}{2}z^{\nu}(\varepsilon_\mu q_\nu-\varepsilon_\nu q_\mu)\,e^{iq.z} $.

In the case of non-perturbative contributions, one of the light quark propagators in Eq.~(\ref{edmn13}), which describes the photon emission at large distances, is replaced with
\begin{align}
\label{edmn21}
S_{\alpha\beta}^{ab}(x) \rightarrow -\frac{1}{4} \big[\bar{q}^a(x) \Gamma_i q^b(x)\big]\big(\Gamma_i\big)_{\alpha\beta},
\end{align}
and in Eq.~(\ref{QCD2}) the remaining four quark propagators are all considered to be full quark propagators. 
Here $\Gamma_i$ represents the full set of Dirac matrices. Once Eq.~(\ref{edmn21}) is inserted into Eq.~(\ref{QCD2}), matrix elements of type $\langle \gamma(q)\vel \bar{q}(x) \Gamma_i q(0) \ver 0\rangle$ and $\langle \gamma(q)\vel \bar{q}(x) \Gamma_i G_{\alpha\beta}q(0) \ver 0\rangle$ appear, which represent the non-perturbative contributions.  We require these matrix elements, which are parameterized in terms of photon wave functions with certain twists, to calculate the non-perturbative contributions (see Ref.~\cite{Ball:2002ps} for the explicit expressions of photon distribution amplitudes (DAs)).  
The QCD side of the correlation function can be acquired in terms of the quark-gluon parameters and the DAs of the photon with the help of Eqs. ~(\ref{QCD2})$-$(\ref{edmn21}) and after Fourier transforms the x-space calculations to momentum space.

 Finally, the Lorentz structure $\eslash \qslash$ is chosen from both the hadronic and the QCD sides, and the coefficients of this structure are matched with each other.  
 Then double Borel transformations and a continuum subtraction are performed. These are used to eliminate the effects of the continuum and higher states.  This is followed by the required QCD light-cone sum rules for the magnetic dipole moments:
\begin{align}
\label{edmn15}
\mu_{\Sigma_{c}^{+}}  =\frac{e^{\frac{m^2_{\Sigma_{c}^{+}}}{M^2}}}{\,\lambda^2_{\Sigma_{c}^{+}}\, m_{\Sigma_{c}^{+}}}\, \Delta_1^{QCD} (M^2,s_0).
\end{align}

The corresponding results for the  $\Delta_1^{QCD}(M^2,s_0)$ function can be found in Appendix. 
As you can see in the equation above, there are two extra parameters: the continuum threshold $s_0$ and the Borel parameters $M^2$. In the numerical analysis section, we will explain how the working intervals of these extra parameters are determined.

\subsection{Magnetic dipole, electric quadrupole and magnetic octupole moments of the $\Lambda_c^+ $ state} 

The required correlation function for the electromagnetic multipole moments of the $\Lambda_c^+ $ state is given by 
\begin{eqnarray} \label{Pc101}
\Pi_{\mu\nu}(p,q)&=&i\int d^4x e^{ip \cdot x} \langle0|T\left\{J_\mu^{\Lambda_c^+}(x)\bar{J}_\nu^{\Lambda_c^+}(0)\right\}|0\rangle _\gamma \, .
\end{eqnarray}
The interpolating current used for $\Lambda_c^+ $ state with
isospin and spin-parity $I(J^P) = 0(3/2^-)$ is as follows:
\begin{align}\label{curpcs2}
J_\mu^{\Lambda_{c}^+}(x)&=\frac{1}{\sqrt{2}}\Big \{\mid  D^{*0} p \rangle \, - \mid  D^{*+} n \rangle  \Big \}
\nonumber\\
& =\frac{1}{\sqrt{2}} \Big \{ \big[\bar u^d(x) \gamma_\mu c^d(x)\big]\big[\varepsilon^{abc} u^{a^T}(x)C\gamma_\alpha d^b(x) \gamma^\alpha \nonumber\\
& \times \gamma_5 u^c(x)\big] + \big[\bar d^d(x) \gamma_\mu c^d(x)\big]  
 \big[\varepsilon^{abc} u^{a^T}(x)C\gamma_\alpha 
 \nonumber\\
& \times d^b(x) 
\gamma^\alpha\gamma_5 d^c(x)\big] \Big\} \,.  
 \end{align}

The correlation function obtained in terms of hadronic parameters is written as
\begin{align}\label{Pc103}
\Pi^{Had}_{\mu\nu}(p,q)&=\frac{\langle0\mid  J_{\mu}^{\Lambda_c^+}(x)\mid
{\Lambda_c^+}(p,s)\rangle}{[p^{2}-m_{{\Lambda_c^+}}^{2}]}
\nonumber\\
& \times  
\langle {\Lambda_c^+}(p,s)\mid
{\Lambda_c^+}(p+q,s)\rangle_\gamma 
\nonumber\\
& \times 
\frac{\langle {\Lambda_c^+}(p+q,s)\mid
\bar{J}_{\nu}^{\Lambda_c^+}(0)\mid 0\rangle}{[(p+q)^{2}-m_{{\Lambda_c^+}}^{2}]}+...
\end{align}
The matrix elements $\langle0\mid J_{\mu}^{\Lambda_c^+}(x)\mid {\Lambda_c^+}(p,s)\rangle$,  $\langle {\Lambda_c^+}(p+q,s)\mid \bar{J}_{\nu}^{\Lambda_c^+}(0)\mid 0\rangle$ and $\langle {\Lambda_c^+}(p)\mid {\Lambda_c^+}(p+q)\rangle_\gamma$ are expressed as follows,
\begin{align}\label{lambdabey}
\langle0\mid J_{\mu}^{\Lambda_c^+}(x)\mid {\Lambda_c^+}(p,s)\rangle&=\lambda_{{\Lambda_c^+}}u_{\mu}(p,s),\\
\langle {\Lambda_c^+}(p+q,s)\mid
\bar{J}_{\nu}^{\Lambda_c^+}(0)\mid 0\rangle &= \lambda_{{\Lambda_c^+}}\bar u_{\nu}(p+q,s), \\
\langle {\Lambda_c^+}(p,s)\mid {\Lambda_c^+}(p+q,s)\rangle_\gamma &=-e\bar
u_{\mu}(p,s)\Bigg[F_{1}(q^2)g_{\mu\nu}\eslash 
\nonumber\\
& 
-
\frac{1}{2m_{{\Lambda_c^+}}} 
\Big[F_{2}(q^2)g_{\mu\nu} 
\nonumber\\
&
+F_{4}(q^2)\frac{q_{\mu}q_{\nu}}{(2m_{{\Lambda_c^+}})^2}\Big]\eslash\qslash
\nonumber\\
&+
F_{3}(q^2)\frac{1}{(2m_{{\Lambda_c^+}})^2}q_{\mu}q_{\nu}\eslash \Bigg]
\nonumber\\
& \times 
u_{\nu}(p+q,s),
\label{matelpar}
\end{align}
where $F_i$'s are the Lorentz invariant form factors and; the $u_{\mu}(p,s)$, $u_{\nu}(p+q,s)$ and $\lambda_{{\Lambda_c^+}}$ are the spinors and residue of the $\Lambda_c^+$ state, respectively.

In principle, we can use Eqs. (\ref{Pc103})$-$(\ref{matelpar}) to get the final expression of the hadronic side of the correlation function, but we run into two problems: not all Lorentz structures are independent, and the correlation function also receives contributions from spin-1/2 particles, which must be eliminated for the calculations to be reliable. The matrix element 
of the current $J_{\mu}$ between spin-1/2 pentaquarks and vacuum is nonzero and is determined as
\begin{equation}\label{spin12}
\langle0\mid J_{\mu}(0)\mid B(p,s=1/2)\rangle=(A  p_{\mu}+B\gamma_{\mu})u(p,s=1/2).
\end{equation}
As is seen the unwanted spin-1/2 contributions are proportional to $\gamma_\mu$ and $p_\mu$.
 By multiplying both sides with $\gamma^\mu$ and using 
 the condition $\gamma^\mu J_\mu = 0$ one can determine the constant A in terms of B.
To obtain only independent structures and to eliminate the spin-1/2 contaminations in the correlation function, we apply the ordering for Dirac matrices as $\gamma_{\mu}\pslash\eslash\qslash\gamma_{\nu}$ and eliminate terms with $\gamma_\mu$ at the beginning, $\gamma_\nu$ at the end and those proportional to $p_\mu$ and $p_\nu$~\cite{Belyaev:1982cd}.  
Applying the above-mentioned manipulations, the correlation function becomes the following 
\begin{align}\label{final phenpart}
\Pi^{Had}_{\mu\nu}(p,q)&=\frac{\lambda_{_{{\Lambda_c^+}}}^{2}}{[(p+q)^{2}-m_{_{{\Lambda_c^+}}}^{2}][p^{2}-m_{_{{\Lambda_c^+}}}^{2}]} 
\nonumber\\
&\times
\Bigg[  g_{\mu\nu}\pslash\eslash\qslash \,F_{1}(q^2) 
-m_{{\Lambda_c^+}}g_{\mu\nu}\eslash\qslash\,F_{2}(q^2)
\nonumber\\
&
-
\frac{F_{3}(q^2)}{4m_{{\Lambda_c^+}}}q_{\mu}q_{\nu}\eslash\qslash
-
\frac{F_{4}(q^2)}{4m_{{\Lambda_c^+}}^3}(\varepsilon.p)q_{\mu}q_{\nu}\pslash\qslash 
\nonumber\\
&
+
\cdots 
\Bigg].
\end{align}
Here, summation over spins of $\Lambda_c$ state is applied as:
\begin{align}\label{raritabela}
\sum_{s}u_{\mu}(p,s)\bar u_{\nu}(p,s)&=-\Big(\pslash+m_{\Lambda_c^+}\Big)\Big[g_{\mu\nu}
-\frac{1}{3}\gamma_{\mu}\gamma_{\nu}
\nonumber\\
& 
-\frac{2\,p_{\mu}p_{\nu}}
{3\,m^{2}_{{\Lambda_c^+}}}+\frac{p_{\mu}\gamma_{\nu}-p_{\nu}\gamma_{\mu}}{3\,m_{{\Lambda_c^+}}}\Big].
\end{align} 
 The final expression obtained for the hadronic representation of the correlation function together with the chosen Lorentz structures is written as follows:

\begin{eqnarray}
\Pi^{Had}_{\mu\nu}(p,q)&=&\Pi_1^{Had}g_{\mu\nu}\pslash\eslash\qslash \,
+\Pi_2^{Had}g_{\mu\nu}\eslash\qslash\,+
\Pi_{3}^{Had}q_{\mu}q_{\nu}\eslash\qslash \,
\nonumber\\
&&
+\Pi_{4}^{Had}(\varepsilon.p)q_{\mu}q_{\nu}\pslash\qslash\,+
...,
\end{eqnarray}
where $ \Pi_1^{Had} $, $ \Pi_2^{Had} $, $ \Pi_3^{Had} $ and $ \Pi_4^{Had} $ are functions of the form factors $ F_1(q^2) $, $ F_2(q^2) $, $ F_3(q^2) $ and  $ F_4(q^2) $, respectively; and dots denote other independent structures.

The form factors of the magnetic dipole, $G_{M}(q^2)$, electric quadrupole, $G_{Q}(q^2)$, and magnetic octupole, $G_{O}(q^2)$ are expressed concerning the form factors $F_{i}(q^2)$ in the following manner~\cite{Weber:1978dh,Nozawa:1990gt,Pascalutsa:2006up,Ramalho:2009vc}:
\begin{eqnarray}
G_{M}(q^2) &=& \left[ F_1(q^2) + F_2(q^2)\right] ( 1+ \frac{4}{5}
\tau ) -\frac{2}{5} \left[ F_3(q^2)  \right]
\nonumber\\
&&+\left[
F_4(q^2)\right] \tau \left( 1 + \tau \right), \nonumber\\
G_{Q}(q^2) &=& \left[ F_1(q^2) -\tau F_2(q^2) \right]  -
\frac{1}{2}\left[ F_3(q^2) -\tau F_4(q^2)
\right] \nonumber\\
&& \times \left( 1+ \tau \right),  \nonumber 
\end{eqnarray}
\begin{eqnarray}
 G_{O}(q^2) &=&
\left[ F_1(q^2) + F_2(q^2)\right] -\frac{1}{2} \left[ F_3(q^2)  +
F_4(q^2)\right]\nonumber\\
&& \times \left( 1 + \tau \right),
\end{eqnarray}
  where $\tau
= -\frac{q^2}{4m^2_{{\Lambda_c^+}}}$. In the static limit, the electromagnetic multipole form factors are given by the form factors $F_i(q^2=0)$ as
\begin{eqnarray}\label{mqo1}
G_{M}(q^2=0)&=&F_{1}(q^2=0)+F_{2}(q^2=0),\nonumber\\
G_{Q}(q^2=0)&=&F_{1}(q^2=0)-\frac{1}{2}F_{3}(q^2=0),\nonumber\\
G_{O}(q^2=0)&=&F_{1}(q^2=0)+F_{2}(q^2=0)-\frac{1}{2}[F_{3}(q^2=0)\nonumber\\
&&+F_{4}(q^2=0)].
\end{eqnarray}
The magnetic dipole ($\mu_{{\Lambda_c^+}}$), electric quadrupole ($Q_{{\Lambda_c^+}}$), and magnetic octupole ($O_{{\Lambda_c^+}}$) moments are written as follows:
 \begin{eqnarray}\label{mqo2}
\mu_{{\Lambda_c^+}}&=&\frac{e}{2m_{{\Lambda_c^+}}}G_{M}(q^2=0),\nonumber\\
Q_{{\Lambda_c^+}}&=&\frac{e}{m_{{\Lambda_c^+}}^2}G_{Q}(q^2=0),\nonumber\\
O_{{\Lambda_c^+}}&=&\frac{e}{2m_{{\Lambda_c^+}}^3}G_{O}(q^2=0).
\end{eqnarray}

The hadronic side of the correlation function is complete when the procedures described above have been applied to it. 
The next step is to determine the QCD side of the correlation function in terms of the QCD parameters, such as the quark-gluon parameters and the photon DAs. The following result is obtained by repeating the steps of the previous subsection:
\begin{align}\label{QCD4}
\Pi_{\mu\nu}^{QCD}(p,q)&= - \frac{i}{2}\varepsilon^{abc} \varepsilon^{a^{\prime}b^{\prime}c^{\prime}}\, \int d^4x \, e^{ip\cdot x} \langle 0\mid \Big\{ 
\nonumber\\
& 
\, \mbox{Tr}\Big[\gamma_\mu S_{c}^{dd^\prime}(x) \gamma_\nu  S_{u}^{d^\prime d}(-x)\Big]  
\mbox{Tr}\Big[\gamma_{\alpha} S_d^{bb^\prime}(x) \gamma_{\beta} \nonumber\\
& \times \widetilde S_{u}^{aa^\prime}(x)\Big]
(\gamma^{\alpha}\gamma_5 S_{u}^{cc^\prime}(x) \gamma_5  \gamma^{\beta})
 \nonumber\\
&     
- \mbox{Tr}\Big[\gamma_\mu S_{c}^{dd^\prime}(x) \gamma_\nu  S_{u}^{d^\prime d}(-x)\Big]   
 \mbox{Tr}\Big[\gamma^{\alpha}\gamma_5 S_{u}^{ca^\prime}(x) \nonumber\\
& \times \gamma_{\beta} \widetilde S_d^{bb^\prime}(x) \gamma_{\alpha}  
 S_{u}^{ac^\prime}(x)  \gamma_5  \gamma^{\beta}\Big]  \nonumber\\
&
  + \mbox{Tr}\Big[\gamma_\mu S_{c}^{dd^\prime}(x) \gamma_\nu  S_{d}^{d^\prime d}(-x)\Big]  
\mbox{Tr}\Big[\gamma_{\alpha} S_d^{bb^\prime}(x) 
\nonumber\\
& \times \gamma_{\beta} \widetilde S_{u}^{aa^\prime}(x)\Big]
(\gamma^{\alpha}\gamma_5 S_{d}^{cc^\prime}(x) \gamma_5  \gamma^{\beta})
 \nonumber\\
&  -  \mbox{Tr}\Big[\gamma_\mu S_{c}^{dd^\prime}(x) \gamma_\nu  S_{d}^{d^\prime d}(-x)\Big]   
 \mbox{Tr}\Big[\gamma^{\alpha}\gamma_5 S_{d}^{cb^\prime}(x) \nonumber\\
& \times \gamma_{\beta} \widetilde S_u^{aa^\prime}(x) \gamma_{\alpha}  
 S_{d}^{bc^\prime}(x)  \gamma_5  \gamma^{\beta}\Big] 
 \Big\}
\mid 0 \rangle _\gamma \,.
\end{align}

The final expression obtained for the QCD representation of the correlation function together with the chosen Lorentz structures is written as follows

\begin{eqnarray}
\Pi^{QCD}_{\mu\nu}(p,q)&=&\Pi_{1}^{QCD}g_{\mu\nu}\pslash\eslash\qslash \,
+\Pi_{2}^{QCD}g_{\mu\nu}\eslash\qslash\,
\nonumber\\
&&+
\Pi_{3}^{QCD}q_{\mu}q_{\nu}\eslash\qslash \,
+\Pi_{4}^{QCD}(\varepsilon.p)q_{\mu}q_{\nu}\pslash\qslash\,+
....,
\end{eqnarray}
where $ \Pi_1^{QCD} $, $ \Pi_2^{QCD} $, $ \Pi_3^{QCD} $ and $ \Pi_4^{QCD} $ are functions of the QCD degrees of freedom and photon DAs parameters. 
%

Both hadronic and quark-gluon parameters have been used to obtain the correlation function. To analyze the magnetic dipole and higher multipole moments, the QCD and hadronic representations of the correlation function are matched through the quark-hadron duality ansatz. By matching the coefficients of the structures $g_{\mu\nu}\pslash\eslash\qslash$, $g_{\mu\nu}\eslash\qslash$, $q_{\mu}q_{\nu}\eslash\qslash$ and  $(\varepsilon.p)q_{\mu}q_{\nu}\pslash\qslash$, respectively for the $F_1$, $F_2$, $F_3$ and $F_4$   we obtain QCD light-cone sum rules for these four invariant form factors. The explicit forms of the expressions obtained for form factors $F_1$, $F_2$, $F_3$, and $F_4$ are given in Appendix \ref{appendb}.

Analytical results are obtained for $\Sigma_c^+$ and $\Lambda_c^+$ states. Performing numerical calculations for these states would be the next step.  
%


\section{Numerical illustrations and discussion of the results} \label{secIII}

The numerical analysis for the magnetic dipole moments of the $\Sigma_c^+$ and $\Lambda_c^+$ states is presented in this section.  The QCD parameters that are used in our calculations are as follows:
  $m_u=m_d=0$, 
$m_c = 1.27\pm 0.02\,$GeV~\cite{ParticleDataGroup:2022pth}, 
$m_{\Sigma_c^{+}}=2792^{+14}_{-5}$ MeV, $m_{\Lambda_c^{+}}=2939.6^{+1.3}_{-1.5}$~MeV
~\cite{ParticleDataGroup:2022pth}, 
$\lambda_{\Sigma_c^{+}} =( 7.98^{+1.98}_{-1.56}) \times 10^{-4}$\,GeV$^6$, 
$\lambda_{\Lambda_c^{+}} =( 8.87^{+1.84}_{-1.59}) \times 10^{-4}$\,GeV$^6$~\cite{Xin:2023gkf},  
$\langle \bar qq\rangle $=$(-0.24\pm 0.01)^3\,$GeV$^3$~\cite{Ioffe:2005ym},   
$m_0^{2} = 0.8 \pm 0.1$~GeV$^2$, $\langle g_s^2G^2\rangle = 0.88~ $GeV$^4$~\cite{Matheus:2006xi}, $\chi=-2.85 \pm 0.5~\mbox{GeV}^{-2}$~\cite{Rohrwild:2007yt} and $f_{3\gamma}=-0.0039~$GeV$^2$~\cite{Ball:2002ps}. 
 One of the important input parameters in the numerical analysis of the calculation of the magnetic dipole and higher multipole moments by the QCD light-cone sum rule method are the photon DAs.
 The photon DAs and the parameters that are used in these DAs are  
 listed in Appendix \ref{appendc}.

In addition to these numerical parameters, which we have already mentioned in the previous section, the sum rules also depend on the auxiliary parameters: Borel mass squared parameter $M^2$ and continuum threshold $s_0$.  Physical measurables such as magnetic dipole and higher multipole moments should be varied as slightly as possible by these extra parameters. Therefore, we are looking for working intervals for these additional parameters such that the results for the magnetic dipole moments are almost independent of these parameters. The standard prescription of the technique used, the Operator Product Expansion (OPE) convergence, and the Pole Contribution (PC) dominance are taken into account in the determination of the working ranges for the parameters $M^2$ and $s_0$. For the characterization of the above restrictions, it is convenient to use the equations below:
\begin{align}
 \mbox{PC} &=\frac{\Delta (M^2,s_0)}{\Delta (M^2,\infty)} \geq  30\%,
\end{align}
and 
\begin{align}
 \mbox{OPE} &=\frac{\Delta^{\mbox{DimN}} (M^2,s_0)}{\Delta (M^2,s_0)}\leq  5\%,
 \end{align}
 where $\Delta^{\mbox{DimN}}(M^2,s_0)=\Delta^{\mbox{Dim(10+11+12)}}(M^2,s_0)$.  
 Based on these restrictions, the working intervals for the auxiliary parameters are as follows:
\begin{align}
 3.0 ~\mbox{GeV}^2 \leq M^2 \leq 5.0 ~\mbox{GeV}^2,\nonumber\\
 11.2 ~\mbox{GeV}^2 \leq s_0 \leq 12.6 ~\mbox{GeV}^2,\nonumber
\end{align}
for the $\Sigma_c^+$ state, and 
\begin{align}
 3.0 ~\mbox{GeV}^2 \leq M^2 \leq 5.0 ~\mbox{GeV}^2,\nonumber\\
 12.3 ~\mbox{GeV}^2 \leq s_0 \leq 13.7 ~\mbox{GeV}^2,\nonumber
\end{align}
for the $\Lambda_c^+$ state. 

Our numerical calculations show that the magnetic dipole moments of these states PC vary within the interval $38\% \leq \mbox{PC} \leq 58\%$ and $33\% \leq \mbox{PC} \leq 57\%$ for the $\Sigma_c^+$ and $\Lambda_c^+$ states, respectively, corresponding to the upper and lower bounds of the Borel mass parameter, by considering these working intervals for the helping parameters. Analyzing the convergence of the OPE, it can be seen that the contribution of the higher twist and higher dimensional terms in the OPE are $3.62\%$ and $3.87\%$ for the $\Sigma_c^+$ and $\Lambda_c^+$ states, respectively, of the total and the series shows good convergence. 
It follows that the constraints imposed by the dominance of PC and the convergence of OPE are satisfied by the working regions determined for $M^2$ and $s_0$.
After the determination of the working intervals of $M^2$ and $s_0$, we now study the dependence of the magnetic dipole moments on $M^2$ for different fixed values of $s_0$.  From Fig.  1 we can see that the magnetic dipole moments show good stability with respect to the variation of $M^2$ in its working interval.
%
\begin{figure}[htp]
\label{Msqfig}
\centering
 \includegraphics[width=0.45\textwidth]{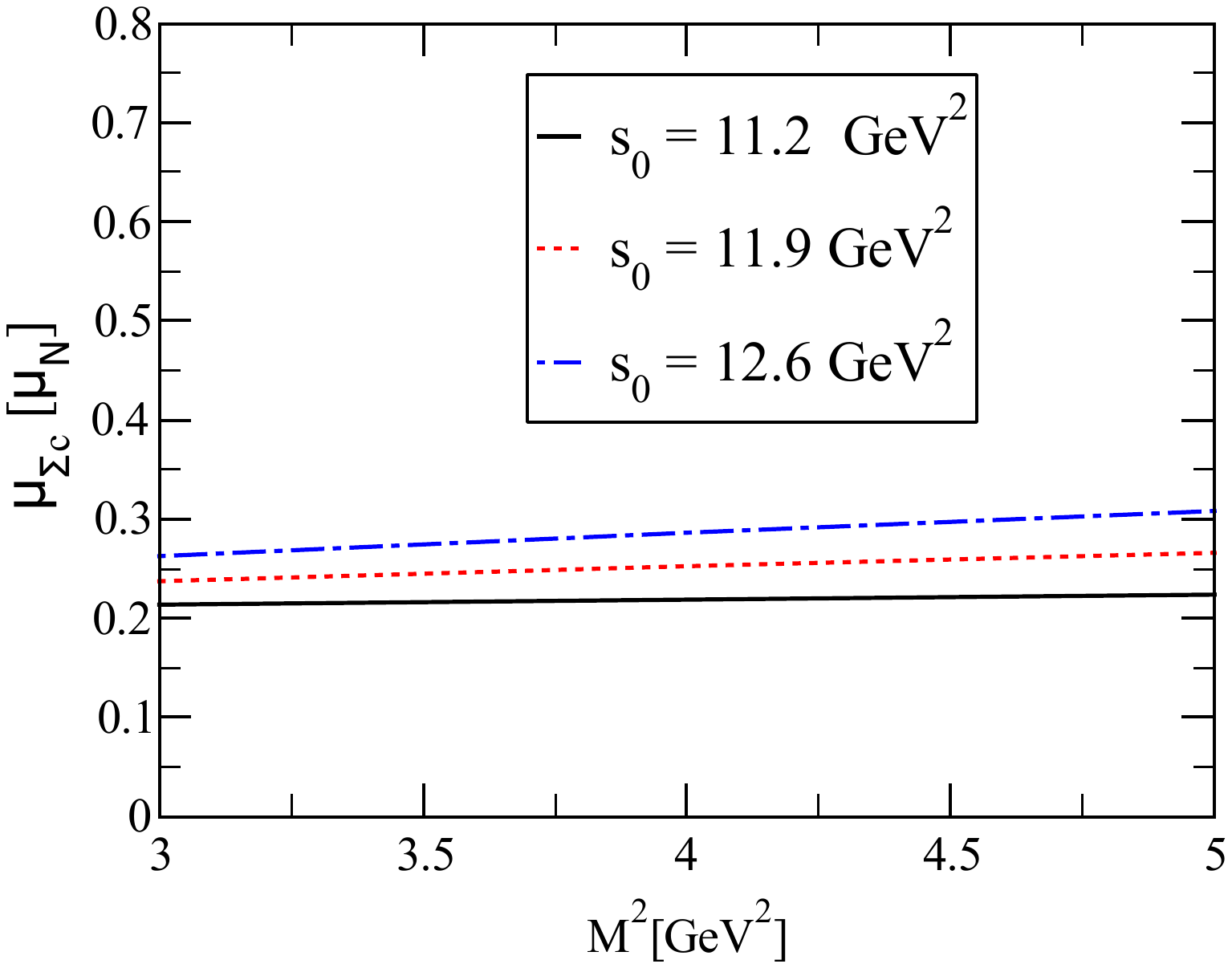}\\
 \vspace{0.5 cm}
 \includegraphics[width=0.45\textwidth]{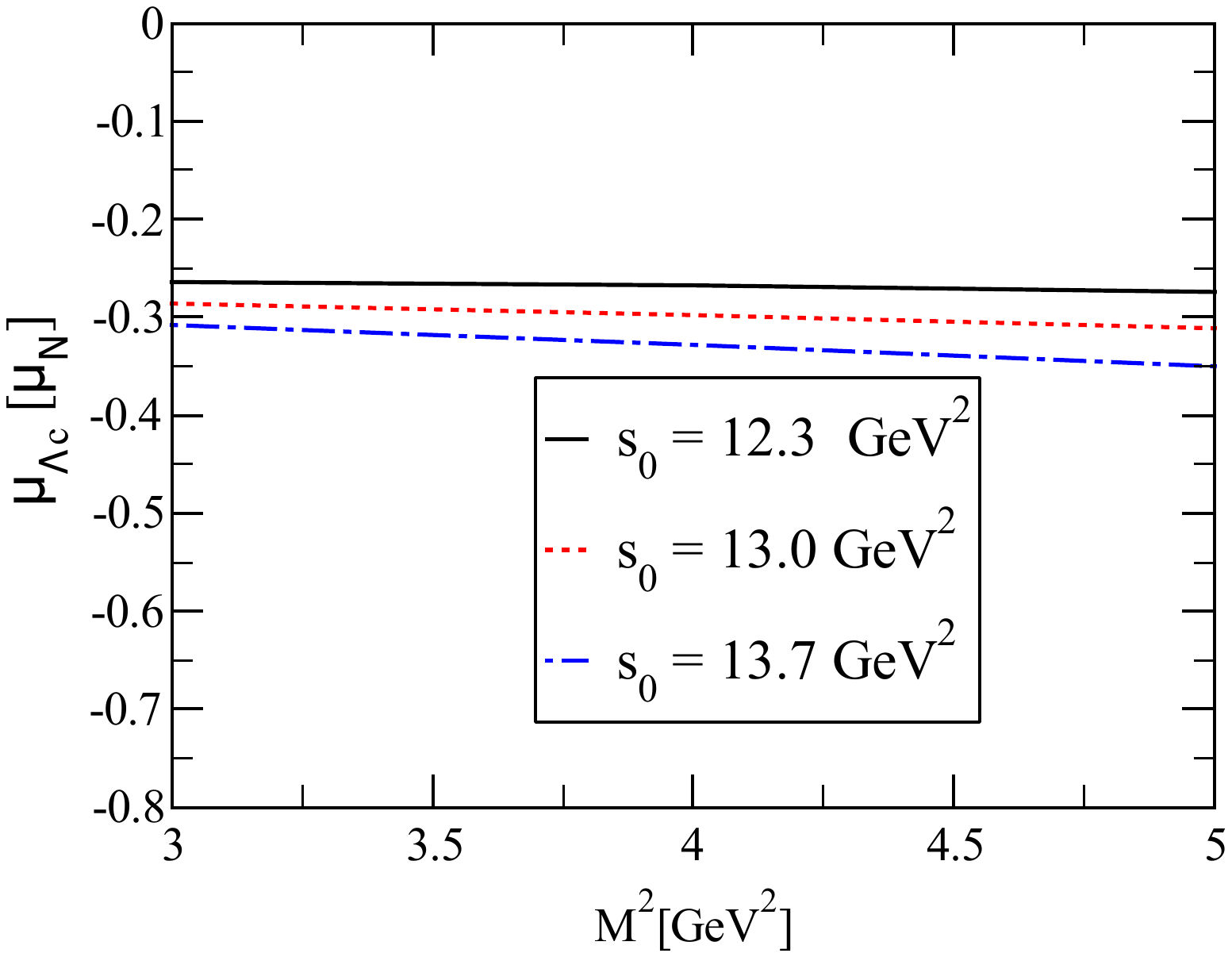}
 \caption{ The dependencies of magnetic dipole moments of the $\Sigma_c^+$ and $\Lambda_c^+$states; on $M^{2}$  at three different values of $s_0$.}
  \end{figure}
%

To give numerical values for the magnetic dipole moments of the $\Sigma_c^+$ and $\Lambda_c^+$ states, we have determined all the necessary parameters. Following our extensive numerical calculations, the magnetic dipole moments, both in their natural unit ($\frac{e}{2 m_{B}}$) and in the nuclear magneton ($\mu_N$), are given as follows: 
\begin{align}
 \mu_{\Sigma_c^{+}}&=0. 78 \pm 0.24 \, \frac{e}{2 m_{\Sigma_c^+}}=0.26 \pm 0.05~\mu_N,\\
 \nonumber\\
 \mu_{\Lambda_c^{+}}&=-0. 97 \pm 0.13 \, \frac{e}{2 m_{\Lambda_c^+}}=-0.31 \pm 0.04~\mu_N.
\end{align}

It should be emphasized here that the numerical computations consider uncertainties in the input parameters, ambiguities entering the photon DAs, as well as uncertainties due to variations of the auxiliary parameters $M^2$ and $s_0$. 

The magnitude of magnetic dipole moments can provide information about their experimental measurability, and the results obtained for magnetic dipole moments show that they are experimentally accessible.  
%
The facilities such as LHCb, Belle II, BESIII and so on may be able to measure magnetic dipole moments of $\Sigma_c^+$ and $\Lambda_c^+$ states with the increased luminosity in future runs and our predictions. 
A comparison of the estimates obtained in this study with the results obtained by different approaches will be a test of the consistency of our results. It is highly recommended to study the magnetic dipole moments of $\Sigma_c^+$ and $\Lambda_c^+$ states with Lattice QCD. In Refs.~\cite{Hall:2014uca,Hall:2016kou}, the authors have calculated the magnetic dipole moment and the mass of the $\Lambda(1405)$ baryon by assuming it as a $\bar K N$ molecular configuration via lattice QCD. The magnetic dipole moments of the $\Sigma_c^+$ and $\Lambda_c^+$ states can be calculated similarly with the help of lattice QCD. Compared with the predicted magnetic dipole moments of lattice QCD, the results of the present study can provide further information for the experimental search for the $\Sigma_c^+$ and $\Lambda_c^+$ states, and the experimental measurements for these magnetic dipole moments could be an important test for the  $DN^{(*)}$ molecular configuration of these states.  Our understanding of their properties and nature will be enhanced by these efforts.

The electric quadrupole ($\mathcal Q_{\Lambda_c^+}$) and the magnetic octupole ($\mathcal O_{\Lambda_c^+}$) of the ${\Lambda_c^+}$ state are also calculated. The predicted results are as follows
 
  \begin{align}
 \mathcal Q_{\Lambda_c^+} &= (0.65 \pm 0.25) \times 10^{-3}~\mbox{fm$^2$},\\
 \nonumber\\
  \mathcal O_{\Lambda_c^+} &=-(0.38 \pm 0.10) \times 10^{-3}~\mbox{fm$^3$}.
 \end{align}

These results are small but non-zero, implying that the $\Lambda_c^+$ state charge distribution is non-spherical.  
Furthermore, the electric quadrupole moment has a positive sign, corresponding to the prolate charge distribution.

\section{summary and concluding remarks}\label{secIV}

Inspired by the controversial and intriguing nature of $\Sigma_{c}(2800)^+$ and $\Lambda_c(2940)^+$ states, the magnetic dipole moments of these states have been calculated in the framework of the QCD light-cone sum rules, using the photon DAs and the assumption that the $\Sigma_{c}(2800)^+$ and $\Lambda_c(2940)^+$ are hadronic molecular states. The obtained result may be useful in determining the exact nature of these states.  To understand the inner structure as well as the geometric shape of the $\Sigma_{c}(2800)^+$ and $\Lambda_c(2940)^+$ states, the confirmation of our predictions with different theoretical models and future experiments can be very helpful. The $\Lambda_c(2940)^+$ state's electric quadrupole and magnetic octupole moments are also extracted. These values indicate that the charge distribution of the $\Lambda_c(2940)^+$ state is non-spherical. We hope that our estimates of the electromagnetic properties of the $\Sigma_{c}(2800)^+$ and $\Lambda_c(2940)^+$ states, together with the results of other theoretical studies of the spectroscopic parameters of these states, will be useful for their search in future experiments and will help us to define the exact internal structures of these states.

  \begin{widetext}

\appendix
\section{Explicit forms of the $\Delta_1^{QCD}(M^2,s_0)$ function } \label{appenda}
In the present appendix, we present explicit expressions of the analytical expressions obtained for the magnetic dipole  moment of the $\Sigma_c^{+}$ state as follows: 
\begin{align}
\Delta_1^{QCD}(M^2,s_0)&= -\frac {P_ 1 P_ 2^2 (e_d + e_u)} {110592 m_c \pi^4}\Bigg[(m_c^4 I[-2, 0] - 
       I[0, 0]) I_ 3[
      h_ {\gamma}] + (-m_c^4 I[-2, 0] + I[0, 0]) \mathbb A[
      u_ 0] + \chi \Big (4 m_c^6 I[-3, 1] \nonumber\\
      &+ 
        m_c^4 (-m_ 0^2 I[-2, 0] + 2 I[-2, 1]) + m_ 0^2 I[0, 0] - 
        2 I[0, 1]\Big) \varphi_ {\gamma}[u_ 0]
      \Bigg]
\nonumber\\
&-  \frac { P_ 1 P_ 2} {3538944 m_c^2 \pi^6}\Bigg[288 (e_d + e_u) \Bigg(
   m_c^8 \Big (I[-4, 2] - 2 I[-3, 1]\Big) + 
    m_c^6 \Big (m_0^2 (2 I[-3, 1] + I[-2, 0]) - 
       4 (I[-3, 2]  \nonumber\\
       &+ I[-2, 1])\Big)+ 
    m_c^4 \Big (3 I[-2, 2] + m_0^2 (I[-2, 1] - I[-1, 0]) - 
       2 I[-1, 1]\Big) +  m_0^2 I[0, 1]\Bigg)\nonumber\\
       &+(e_d + e_u)\Bigg (-39 m_c^4 (m_c^4 I[-4, 2] - 
          I[-2, 2]) I_ 2[\mathcal S] - 
       12 m_c^4  \Big (7 m_c^4 I[-4, 2] - 13 m_c^2 I[-3, 2]
       + 
          6 I[-2, 2]\Big) \nonumber\\
       & \times I_ 3[
         h_ {\gamma}] +6 \Big (14 m_c^8 I[-4, 2] - 13 m_c^6 I[-3, 2] -           I[0, 2]\Big) \mathbb A[u_ 0] - 
       8 \chi \Big (3 m_c^{10} I[-5, 3] + 5 m_c^8 I[-4, 3]\nonumber\\
       &- 
           2 m_c^4 I[-2, 3] + 6 I[0, 3]\Big) \varphi_ {\gamma}[
          u_ 0]\Bigg) - 
    f_ {3\gamma} \pi^2  \Bigg (52 (e_d + 2 e_u)  (m_c^4 I[-2, 1] + 
           I[0, 1]) I_ 1[\mathcal V] - 
        96 (e_d - 2 e_u)\nonumber\\
        & \times (4 m_c^6 I[-3, 1] - m_ 0^2 m_c^4 I[-2, 0] + 
            m_ 0^2 I[0, 0] - 4 I[0, 1]) \psi^a[u_ 0]\Bigg)\Bigg]\nonumber\\
      & +\frac { P_ 2^2} {18432 m_c \pi^4} \Bigg[18(e_d + e_u) \Bigg(
   4 m_c^8 (I[-4, 2] - 2 I[-3, 1]) + 
    8 m_c^6 \Big (m_ 0^2 (2 I[-3, 1] + I[-2, 0]) - 
       2 (I[-3, 2] 
       \nonumber\\
       &+ I[-2, 1])\Big)- 
    m_c^4 \Big (m_ 0^4 I[-2, 0] - 12 I[-2, 2] - 
       8 m_ 0^2 (I[-2, 1] - I[-1, 0]) + 8 I[-1, 1]\Big) + 
    m_ 0^4 I[0, 0]\nonumber\\
    &+ 8 m_ 0^2 I[0, 1]\Bigg)
     +(e_d+e_u)\Bigg(
    9 m_c^4 \Big (m_c^4 I[-4, 2] - m_c^2 (m_0^2 I[-3, 1] + 2 I[-3, 2]) - 
    m_0^2 I[-2, 1] + I[-2, 2]\Big)\nonumber\\
    & \times I_3[h_{\gamma}] - 
 9 \Big (2 m_c^8 I[-4, 2] - 2 m_c^6 (m_0^2 I[-3, 1] + I[-3, 2]) - 
     m_0^2 m_c^4 I[-2, 1] + m_ 0^2 I[0, 1]\Big) \mathbb A[u_ 0] \nonumber\\
     &+ 
 9  \Big(m_c^8 I[-4, 2] - m_0^2 m_c^6 I[-3, 1] - m_c^4 I[-2, 2] + 
    m_ 0^2 I[0,1]\Big) I_2[\mathcal S] + \chi \Big(m_c^8 (4 m_c^2 I[-5,3] - 9 m_0^2 I[-4, 2] \nonumber\\
    &+ 6 I[-4, 3]) + 
     9 m_0^2 m_c^6 I[-3, 2] - 2 m_c^4 I[-2, 3] + 
     8 I[0, 3]\Big) \varphi_{\gamma}[u_ 0]
        \Bigg)
        +6 f_{3 \gamma} \pi^2 \Bigg ((e_d + 2 e_u)  (m_c^4 I[-2, 1]\nonumber\\
        &+ 
      I[0, 1]) I_ 1[\mathcal V] - 
   4 (e_d - 2 e_u) \Big (2 m_c^6 I[-3, 1] - m_ 0^2 m_c^4 I[-2, 0] + 
       m_ 0^2 I[0, 0] - 2 I[0, 1]\Big) \psi^a[u_ 0]\Bigg)\Bigg]\nonumber\\
    &+\frac { P_ 2} {491520 m_c^2 \pi^6}\Bigg[40(e_d + e_u)
  \Bigg( m_c^{12} \Big (3 I[-6, 4] + 4 I[-5, 3]\Big) - 
    3 m_c^{10} \Big (m_ 0^2 (4 I[-5, 3] - 3 I[-4, 2]) \nonumber\\
       &+ 
       4 (I[-5, 4] - I[-4, 3])\Big) - 
    3 m_c^8 \Big (-5 I[-4, 4] + 6 m_ 0^2 (I[-4, 3] + I[-3, 2]) - 
       4 I[-3, 3]\Big) \nonumber\\
    & + 
    m_c^6 \Big (-6 I[-3, 4]+ m_ 0^2 (-6 I[-3, 3] + 9 I[-2, 2]) + 
       4 I[-2, 3]\Big) + 32 m_c^2 I[0, 3] - 36 m_ 0^2 I[0, 3]\Bigg)\nonumber\\
       &+ (e_u+e_d)\Bigg(
       -5 m_c^8 \Big ( (m_c^4 I[-6, 4] - 
       3 I[-4, 4]) I_ 2[\mathcal S] + (m_c^4 I[-6, 4] - 
       3 m_c^2 I[-5, 4] + 3 I[-4, 4]) I_ 6[h_ {\gamma}]\Big) \nonumber\\
       &+ 
 5 m_c^6 (-2 I_ 2[\mathcal S] + I_ 3[h_ {\gamma}]) I[-3, 4] + 
 m_c^6  \big (2 m_c^6 I[-6, 4] - 3 m_c^4 I[-5, 4] + 
     I[-3, 4]\big)\mathbb A[u_ 0] \nonumber\\
     &+ 
 2 \chi \Big (m_c^6 (m_c^6 I[-6, 5] + 3 m_c^4 I[-5, 5] + 
        3 m_c^2 I[-4, 5] + I[-3, 5]) + 
     8 I[0, 5]\Big) \varphi_ {\gamma}[u_ 0]
       \Bigg)\nonumber
               \end{align}
               \begin{align}
               &
               +f_ {3\gamma} \pi^2 \Bigg (-(e_d + 2 e_u)  \Big (4 m_c^8 I[-4, 3] + 
      m_c^6 (-3 m_ 0^2 I[-3, 2] + 4 I[-3, 3]) + 3 m_ 0^2 I[0, 2] + 
      8 I[0, 3]\Big) I_ 1[\mathcal V] \nonumber\\
      &+ 
   8 (e_d - 2 e_u) \Big (2 m_c^{10} I[-5, 3] + 
       3 m_ 0^2 m_c^8 I[-4, 2] - 2 m_c^6 I[-3, 3] - 
       3 m_ 0^2 I[0, 2]\Big)\psi^a[u_ 0]\Bigg)
                \Bigg],
               \end{align}
where $P_1 =\langle g_s^2 G^2\rangle$ and  $P_2 =\langle \bar q q \rangle$  are gluon and light-quark condensates, respectively. It should also be noted that in the $\Delta_1^{QCD}(M^2,s_0)$ function we have, for simplicity, presented only those terms that make significant contributions to the numerical values of the magnetic dipole moment and have ignored many higher dimensional terms, although they are included in the numerical calculations.
The functions~$I[n,m]$, $I_1[\mathcal{F}]$,~$I_2[\mathcal{F}]$~and~$I_3[\mathcal{F}]$
are
defined as:
\begin{align}
 I[n,m]&= \int_{m_c^2}^{s_0} ds \int_{m_c^2}^s dl~ e^{-s/M^2}\,l^n~(s-l)^m ,\nonumber\\
I_1[\mathcal{F}]&=\int D_{\alpha_i} \int_0^1 dv~ \mathcal{F}(\alpha_{\bar q},\alpha_q,\alpha_g)
 \delta^{\prime}(\alpha_{\bar q}+ v \alpha_g-u_0),\nonumber\\
 I_2[\mathcal{F}]&=\int D_{\alpha_i} \int_0^1 dv~ \mathcal{F}(\alpha_{\bar q},\alpha_q,\alpha_g)
 \delta(\alpha_{\bar q}+ v \alpha_g-u_0),\nonumber\\
 I_3[\mathcal{F}]&= \int_0^1 du \, \mathcal{F}(u), 
 \end{align}
 where $\mathcal{F}$ represents the corresponding photon DAs and The measure ${\cal D} \alpha_i$ is defined as

\begin{eqnarray*}
\label{nolabel05}
\int {\cal D} \alpha_i = \int_0^1 d \alpha_{\bar q} \int_0^1 d
\alpha_q \int_0^1 d \alpha_g \delta^{\prime}(1-\alpha_{\bar
q}-\alpha_q-\alpha_g).~\nonumber\\
\end{eqnarray*}
 

\section{  Explicit forms of the sum rules for $F_i$ form factors } \label{appendb}
In the present appendix, we present explicit expressions of the analytical expressions obtained for the magnetic dipole, electric quadrupole, and magnetic octupole moments of the $\Lambda_c^{+}$ state. These multipole moments are written as follows:
 \begin{eqnarray}
\mu_{{\Lambda_c^+}}&=&\frac{e}{2m_{{\Lambda_c^+}}}F_{1}+F_{2},\nonumber\\
Q_{{\Lambda_c^+}}&=&\frac{e}{m_{{\Lambda_c^+}}^2}F_{1}-\frac{1}{2}F_{3},\nonumber\\
O_{{\Lambda_c^+}}&=&\frac{e}{2m_{{\Lambda_c^+}}^3}F_{1}+F_{2}-\frac{1}{2}[F_{3}+F_{4}],
\end{eqnarray}
where
\begin{align}\label{sonF1}
  F_1&=-\frac{e^{\frac{m^2_{\Lambda_{c}^{+}}}{M^2}}}{\,\lambda^2_{\Lambda_{c}^{+}}}\Bigg\{-\frac { P_ 1 P_ 2^2 (e_d + e_u)} {13824 m_c^2 \pi^4}\Big[
   I[0, 0] - 2 m_c^4 I[2, 0] + m_c^6 I[3, 0]\Big]\nonumber\\
   & +\frac {P_ 1 P_ 2 (e_d + 
     e_u)} {589824 m_c \pi^6} \Bigg[ \Big (m_c^6 I[-3, 1] + 
       2 m_c^4 I[-2, 1] + m_c^2 I[-1, 1] + 
       4 I[0, 1]\Big) I_ 1[\mathcal S] - \Big (m_c^6 I[-3, 1] + 
       2 m_c^4 I[-2, 1] \nonumber\\
       &+ m_c^2 I[-1, 1] + 
       4 I[0, 1]\Big) I_ 1[\mathcal {\tilde S}] - 
    20 \Big (4 m_c^6 I[-3, 1] + m_ 0^2 I[0, 0] + 
        m_c^4 (4 I[-2, 1] - m_ 0^2 I[2, 0])\Big)\Bigg]\nonumber\\
       & +\frac {P_ 2^2 (e_d + e_u) } {2048 m_c^2 \pi^4}\Bigg[ 
   4 m_c^{10} I[-5, 2] - 8 m_ 0^2 m_c^8 I[-4, 1] - m_ 0^4 I[0, 0] + 
    8 m_ 0^2 I[0, 1] + 
    m_c^6 \big (-4 I[-3, 2] + m_ 0^4 I[3, 0]\big)\Bigg]\nonumber\\
    & - \frac {P_ 1 (e_d + e_u)} {1769472 m_c^2 \pi^8}\Bigg[ 
    6 m_c^{12} I[-6, 3] + 8 m_c^{10} I[-5, 3] + 3 m_c^8 I[-4, 3] + 
     6 m_c^6 I[-3, 3] + 5 m_c^4 I[-2, 3] + 28 I[0, 3]\Bigg]\nonumber
\end{align}

\begin{align}
& - \frac { 
   P_ 2 (e_d + e_u)} {12288 m_c \pi^6}\Bigg[ 
    m_c^4 \Big (4 m_c^6 I[-5, 3] + 
        m_c^4 \big (-9 m_ 0^2 I[-4, 2] + 12 I[-4, 3]\big) + 
        6 m_c^2 \big (3 m_ 0^2 I[-3, 2] + 2 I[-3, 3]\big) \nonumber\\
        &- 
        9 m_ 0^2 I[-2, 2] + 4 I[-2, 3]\Big) + 32 I[0, 3] \Bigg]\nonumber\\
& - \frac {(e_d + 
      e_u)} {1638400 m_c^2 \pi^8}\Bigg[ m_c^{16} I[-8, 5] - 
       10 m_c^{12} I[-6, 5] - 20 m_c^{10} I[-5, 5] - 
       15 mc^8 I[-4, 5] - 4 mc^6 I[-3, 5] - 48 I[0, 5]\Bigg]\Bigg\},\\
 F_2&= \frac{ m_{\Lambda_c^+} e^{\frac{m^2_{\Lambda_{c}^{+}}}{M^2}}}{\,\lambda^2_{\Lambda_{c}^{+}}}\Bigg\{ \frac {P_ 1 P_ 2 (e_d + e_u)} {36864 m_c^2 \pi^6}\Bigg[ 
   2 m_c^8 \big (I[-4, 2] - I[-3, 1]\big) + 
    m_c^4 \big (4 I[-2, 2] - m_ 0^2 I[-1, 0] - 2 I[-1, 1]\big) + 
    m_ 0^2 I[0, 1] \nonumber\\
    &+ 
    m_c^6 \Big (-6 I[-3, 2] - 4 I[-2, 1] + 
        m_ 0^2 \big (I[-3, 1] + I[2, 0]\big)\Big)  \Bigg]\nonumber\\
     & + \frac {P_ 2^2 (e_d + e_u)} {1024 m_c^2 \pi^4}\Bigg[ 
    4 m_c^8 (I[-4, 2] - 2 I[-3, 1]) + m_ 0^4 I[0, 0] + 
     8 m_ 0^2 I[0, 1] - 
     m_c^4 \Big (-12 I[-2, 2] - 8 m_ 0^2 (I[-2, 1]  \nonumber\\
     &- I[-1, 0])+ 
        8 I[-1, 1] + m_ 0^4 I[2, 0]\Big) + 
     8 m_c^6 \Big (-2 (I[-3, 2] + I[-2, 1]) + 
         m_ 0^2 (2 I[-3, 1] + I[2, 0])\Big) \Bigg]\nonumber\\
     &+\frac {P_ 2 (e_d + e_u)} {12288 m_c^2 \pi^6}\Bigg[ 
   2 m_c^{12} \big (I[-6, 4] + 2 I[-5, 3]\big) - 
    3 m_c^{10}\Big (3 I[-5, 4] + 
       m_ 0^2 \big (5 I[-5, 3] - 3 I[-4, 2]\big) - 4 I[-4, 3]\Big)
       \nonumber\\
       &- 
    6 m_c^8 \Big (m_ 0^2 (4 I[-4, 3] + 3 I[-3, 2]) - 
       2 (I[-4, 4] + I[-3, 3])\Big) + 
    m_c^6 \Big (-5 I[-3, 4] - 9 m_ 0^2 (I[-3, 3] - I[-2, 2]) \nonumber\\
       &+ 
       4 I[-2, 3]\Big) - 48 m_ 0^2 I[0, 3] + 32 m_c^2 I[0, 3] \Bigg]\Bigg\},\\
 F_3&=\frac{ 4 m_{\Lambda_c^+} e^{\frac{m^2_{\Lambda_{c}^{+}}}{M^2}}}{\,\lambda^2_{\Lambda_{c}^{+}}}\Bigg\{-\frac {P_ 1 P_ 2 (e_d + e_u)} {55296 m_c^2 \pi^6} \Bigg[ 
   m_c^{10} (I[-5, 1] + I[4, 0]) - 3 m_c^8 (I[-4, 1] + I[3, 0]) + 
    3 m_c^6 (-3 I[-3, 1] + I[2, 0] \nonumber\\
       & - m_ 0^2 I[3, 0])- 
    m_c^4 (5 I[-2, 1] + I[-1, 0] - 3 m_ 0^2 I[2, 0]) - 
    16 I[0, 1] \Bigg]\nonumber\\
   & -\frac {P_ 2 (e_d + e_u)} {12288 m_c^2 \pi^6} \Bigg[ 
   m_c^6 \Bigg (3 m_c^8 (I[-7, 3] - I[-6, 2]) + 
       m_c^6 \Big (-6 m_ 0^2 (I[-6, 2] + I[-5, 1]) + 
          4 (I[-6, 3] + 3 I[-5, 2])\Big) \nonumber\\
       &- 
       6 m_c^4 \Big (I[-5, 3] + 3 m_ 0^2 (-2 I[-5, 2] + I[-4, 1]) + 
          3 I[-4, 2]\Big) - 
       6 m_c^2 \Big (3 m_ 0^2 (3 I[-4, 2] + I[-3, 1]) + 
          2 (I[-4, 3] \nonumber\\
       &- I[-3, 2])\Big) - 5 I[-3, 3] + 
       6 m_ 0^2 (4 I[-3, 2] - I[-2, 1]) - 3 I[-2, 2]\Bigg) - 
    48 m_ 0^2 m_c^2 I[0, 1] - 16 I[0, 3] \Bigg]\Bigg\},
\end{align}
and,  
\begin{align}\label{sonF4}
 F_4&=\frac{ 4 m^3_{\Lambda_c^+} e^{\frac{m^2_{\Lambda_{c}^{+}}}{M^2}}}{\,\lambda^2_{\Lambda_{c}^{+}}}\Bigg\{\frac {P_ 1 P_ 2 (e_d + e_u)} {27648 \pi^6} \Bigg[-I[2, 0] + 
    3 m_c^2 I[3, 0] - 3 m_c^4 I[4, 0] + m_c^6 I[5, 0] \Bigg]\nonumber\\
       &
       +\frac {P_ 2 (e_d + e_u)} {2048 m_c^2 \pi^6} \Bigg[
   m_c^6 \Bigg (-m_c^8 I[-7, 2] + 
       2 m_c^6 (m_ 0^2 I[-6, 1] + 2 I[-6, 2]) + 
       6 m_c^4 (m_ 0^2 I[-5, 1] - I[-5, 2])+ 
       m_c^2 (6 m_ 0^2  
       \nonumber\\
       & \times I[-4, 1] + 4 I[-4, 2]) + 2 m_ 0^2 I[-3, 1] - 
       I[-3, 2]\Bigg) + 16 m_ 0^2 I[0, 1]\Bigg]\Bigg\}.
\end{align}
Definitions of $P_1$, $P_2$, $I[n,m]$, $I_1[\mathcal{F}]$, $I_2[\mathcal{F}]$, and $I_3[\mathcal{F}]$ in Eqs. (\ref{sonF1})-(\ref{sonF4}) are given at end of Appendix \ref{appenda}.


\section{ Distribution Amplitudes of the photon } \label{appendc}
In this appendix, the matrix elements $\langle \gamma(q)\vel \bar{q}(x) \Gamma_i q(0) \ver 0\rangle$  
and $\langle \gamma(q)\vel \bar{q}(x) \Gamma_i G_{\mu\nu}q(0) \ver 0\rangle$ associated with the photon DAs are presented as follows \cite{Ball:2002ps}:
\begin{eqnarray*}
\label{esbs14}
&&\langle \gamma(q) \vert  \bar q(x) \gamma_\mu q(0) \vert 0 \rangle
= e_q f_{3 \gamma} \left(\varepsilon_\mu - q_\mu \frac{\varepsilon
x}{q x} \right) \int_0^1 du e^{i \bar u q x} \psi^v(u)
\nonumber \\
&&\langle \gamma(q) \vert \bar q(x) \gamma_\mu \gamma_5 q(0) \vert 0
\rangle  = - \frac{1}{4} e_q f_{3 \gamma} \epsilon_{\mu \nu \alpha
\beta } \varepsilon^\nu q^\alpha x^\beta \int_0^1 du e^{i \bar u q
x} \psi^a(u)
\nonumber \\
&&\langle \gamma(q) \vert  \bar q(x) \sigma_{\mu \nu} q(0) \vert  0
\rangle  = -i e_q \langle \bar q q \rangle (\varepsilon_\mu q_\nu - \varepsilon_\nu
q_\mu) \int_0^1 du e^{i \bar u qx} \left(\chi \varphi_\gamma(u) +
\frac{x^2}{16} \mathbb{A}  (u) \right) \nonumber \\ 
&&-\frac{i}{2(qx)}  e_q \bar qq \left[x_\nu \left(\varepsilon_\mu - q_\mu
\frac{\varepsilon x}{qx}\right) - x_\mu \left(\varepsilon_\nu -
q_\nu \frac{\varepsilon x}{q x}\right) \right] \int_0^1 du e^{i \bar
u q x} h_\gamma(u)
\nonumber \\
&&\langle \gamma(q) | \bar q(x) g_s G_{\mu \nu} (v x) q(0) \vert 0
\rangle = -i e_q \langle \bar q q \rangle \left(\varepsilon_\mu q_\nu - \varepsilon_\nu
q_\mu \right) \int {\cal D}\alpha_i e^{i (\alpha_{\bar q} + v
\alpha_g) q x} {\cal S}(\alpha_i)
\nonumber \\
&&\langle \gamma(q) | \bar q(x) g_s \tilde G_{\mu \nu}(v
x) i \gamma_5  q(0) \vert 0 \rangle = -i e_q \langle \bar q q \rangle \left(\varepsilon_\mu q_\nu -
\varepsilon_\nu q_\mu \right) \int {\cal D}\alpha_i e^{i
(\alpha_{\bar q} + v \alpha_g) q x} \tilde {\cal S}(\alpha_i)
\nonumber \\
&&\langle \gamma(q) \vert \bar q(x) g_s \tilde G_{\mu \nu}(v x)
\gamma_\alpha \gamma_5 q(0) \vert 0 \rangle = e_q f_{3 \gamma}
q_\alpha (\varepsilon_\mu q_\nu - \varepsilon_\nu q_\mu) \int {\cal
D}\alpha_i e^{i (\alpha_{\bar q} + v \alpha_g) q x} {\cal
A}(\alpha_i)
\nonumber \\
&&\langle \gamma(q) \vert \bar q(x) g_s G_{\mu \nu}(v x) i
\gamma_\alpha q(0) \vert 0 \rangle = e_q f_{3 \gamma} q_\alpha
(\varepsilon_\mu q_\nu - \varepsilon_\nu q_\mu) \int {\cal
D}\alpha_i e^{i (\alpha_{\bar q} + v \alpha_g) q x} {\cal
V}(\alpha_i) \nonumber\\
&& \langle \gamma(q) \vert \bar q(x)
\sigma_{\alpha \beta} g_s G_{\mu \nu}(v x) q(0) \vert 0 \rangle  =
e_q \langle \bar q q \rangle \left\{
        \left[\left(\varepsilon_\mu - q_\mu \frac{\varepsilon x}{q x}\right)\left(g_{\alpha \nu} -
        \frac{1}{qx} (q_\alpha x_\nu + q_\nu x_\alpha)\right) \right. \right. q_\beta
\nonumber \\
 && -
         \left(\varepsilon_\mu - q_\mu \frac{\varepsilon x}{q x}\right)\left(g_{\beta \nu} -
        \frac{1}{qx} (q_\beta x_\nu + q_\nu x_\beta)\right) q_\alpha
-
         \left(\varepsilon_\nu - q_\nu \frac{\varepsilon x}{q x}\right)\left(g_{\alpha \mu} -
        \frac{1}{qx} (q_\alpha x_\mu + q_\mu x_\alpha)\right) q_\beta
\nonumber \\
 &&+
         \left. \left(\varepsilon_\nu - q_\nu \frac{\varepsilon x}{q.x}\right)\left( g_{\beta \mu} -
        \frac{1}{qx} (q_\beta x_\mu + q_\mu x_\beta)\right) q_\alpha \right]
   \int {\cal D}\alpha_i e^{i (\alpha_{\bar q} + v \alpha_g) qx} {\cal T}_1(\alpha_i)
\nonumber \\
&&+
        \left[\left(\varepsilon_\alpha - q_\alpha \frac{\varepsilon x}{qx}\right)
        \left(g_{\mu \beta} - \frac{1}{qx}(q_\mu x_\beta + q_\beta x_\mu)\right) \right. q_\nu
\nonumber \\ &&-
         \left(\varepsilon_\alpha - q_\alpha \frac{\varepsilon x}{qx}\right)
        \left(g_{\nu \beta} - \frac{1}{qx}(q_\nu x_\beta + q_\beta x_\nu)\right)  q_\mu
\nonumber \\ 
 && -
         \left(\varepsilon_\beta - q_\beta \frac{\varepsilon x}{qx}\right)
        \left(g_{\mu \alpha} - \frac{1}{qx}(q_\mu x_\alpha + q_\alpha x_\mu)\right) q_\nu
\nonumber \\
 &&+
         \left. \left(\varepsilon_\beta - q_\beta \frac{\varepsilon x}{qx}\right)
        \left(g_{\nu \alpha} - \frac{1}{qx}(q_\nu x_\alpha + q_\alpha x_\nu) \right) q_\mu
        \right]      
    \int {\cal D} \alpha_i e^{i (\alpha_{\bar q} + v \alpha_g) qx} {\cal T}_2(\alpha_i)
\nonumber \\
&&+\frac{1}{qx} (q_\mu x_\nu - q_\nu x_\mu)
        (\varepsilon_\alpha q_\beta - \varepsilon_\beta q_\alpha)
    \int {\cal D} \alpha_i e^{i (\alpha_{\bar q} + v \alpha_g) qx} {\cal T}_3(\alpha_i)
\nonumber \\ &&+
        \left. \frac{1}{qx} (q_\alpha x_\beta - q_\beta x_\alpha)
        (\varepsilon_\mu q_\nu - \varepsilon_\nu q_\mu)
    \int {\cal D} \alpha_i e^{i (\alpha_{\bar q} + v \alpha_g) qx} {\cal T}_4(\alpha_i)
                        \right\}~,
\end{eqnarray*}
where $\varphi_\gamma(u)$ is the DA of leading twist-2, $\psi^v(u)$,
$\psi^a(u)$, ${\cal A}(\alpha_i)$ and ${\cal V}(\alpha_i)$, are the twist-3 amplitudes, and
$h_\gamma(u)$, $\mathbb{A}(u)$, ${\cal S}(\alpha_i)$, ${\cal{\tilde S}}(\alpha_i)$, ${\cal T}_1(\alpha_i)$, ${\cal T}_2(\alpha_i)$, ${\cal T}_3(\alpha_i)$ 
and ${\cal T}_4(\alpha_i)$ are the
twist-4 photon DAs.

The expressions of the DAs that are entered into the matrix elements above are  as follows:
\begin{eqnarray}
\varphi_\gamma(u) &=& 6 u \bar u \left( 1 + \varphi_2(\mu)
C_2^{\frac{3}{2}}(u - \bar u) \right),
\nonumber \\
\psi^v(u) &=& 3 \left(3 (2 u - 1)^2 -1 \right)+\frac{3}{64} \left(15
w^V_\gamma - 5 w^A_\gamma\right)
                        \left(3 - 30 (2 u - 1)^2 + 35 (2 u -1)^4
                        \right),
\nonumber \\
\psi^a(u) &=& \left(1- (2 u -1)^2\right)\left(5 (2 u -1)^2 -1\right)
\frac{5}{2}
    \left(1 + \frac{9}{16} w^V_\gamma - \frac{3}{16} w^A_\gamma
    \right),
\nonumber \\
h_\gamma(u) &=& - 10 \left(1 + 2 \kappa^+\right) C_2^{\frac{1}{2}}(u
- \bar u),
\nonumber \\
\mathbb{A}(u) &=& 40 u^2 \bar u^2 \left(3 \kappa - \kappa^+
+1\right)  +
        8 (\zeta_2^+ - 3 \zeta_2) \left[u \bar u (2 + 13 u \bar u) \right.
\nonumber \\ && + \left.
                2 u^3 (10 -15 u + 6 u^2) \ln(u) + 2 \bar u^3 (10 - 15 \bar u + 6 \bar u^2)
        \ln(\bar u) \right],
        \nonumber \\
        {\cal A}(\alpha_i) &=& 360 \alpha_q \alpha_{\bar q} \alpha_g^2
        \left(1 + w^A_\gamma \frac{1}{2} (7 \alpha_g - 3)\right),
\nonumber \\
{\cal V}(\alpha_i) &=& 540 w^V_\gamma (\alpha_q - \alpha_{\bar q})
\alpha_q \alpha_{\bar q}
                \alpha_g^2,
\nonumber \\
{\cal T}_1(\alpha_i) &=& -120 (3 \zeta_2 + \zeta_2^+)(\alpha_{\bar
q} - \alpha_q)
        \alpha_{\bar q} \alpha_q \alpha_g,
\nonumber \\
{\cal T}_2(\alpha_i) &=& 30 \alpha_g^2 (\alpha_{\bar q} - \alpha_q)
    \left((\kappa - \kappa^+) + (\zeta_1 - \zeta_1^+)(1 - 2\alpha_g) +
    \zeta_2 (3 - 4 \alpha_g)\right),
\nonumber \\
{\cal T}_3(\alpha_i) &=& - 120 (3 \zeta_2 - \zeta_2^+)(\alpha_{\bar
q} -\alpha_q)
        \alpha_{\bar q} \alpha_q \alpha_g,
\nonumber \\
{\cal T}_4(\alpha_i) &=& 30 \alpha_g^2 (\alpha_{\bar q} - \alpha_q)
    \left((\kappa + \kappa^+) + (\zeta_1 + \zeta_1^+)(1 - 2\alpha_g) +
    \zeta_2 (3 - 4 \alpha_g)\right),\nonumber \\
{\cal S}(\alpha_i) &=& 30\alpha_g^2\{(\kappa +
\kappa^+)(1-\alpha_g)+(\zeta_1 + \zeta_1^+)(1 - \alpha_g)(1 -
2\alpha_g)\nonumber +\zeta_2[3 (\alpha_{\bar q} - \alpha_q)^2-\alpha_g(1 - \alpha_g)]\},\nonumber \\
\tilde {\cal S}(\alpha_i) &=&-30\alpha_g^2\{(\kappa -\kappa^+)(1-\alpha_g)+(\zeta_1 - \zeta_1^+)(1 - \alpha_g)(1 -
2\alpha_g)
+\zeta_2 [3 (\alpha_{\bar q} -\alpha_q)^2-\alpha_g(1 - \alpha_g)]\}.
\end{eqnarray}

The numerical values of the parameters used in the DAs are $\varphi_2(1~GeV) = 0$, 
$w^V_\gamma = 3.8 \pm 1.8$, $w^A_\gamma = -2.1 \pm 1.0$, $\kappa = 0.2$, $\kappa^+ = 0$, $\zeta_1 = 0.4$, and $\zeta_2 = 0.3$.

\end{widetext}

\bibliographystyle{elsarticle-num}
\bibliography{Singlycharm.bib}

\end{document}